\pdfoutput=1
\documentclass[aps,pra,showpacs,%
  reprint,
  ]{revtex4-1}
\usepackage{graphicx}
\usepackage{natbib}
\usepackage{amsmath}
\usepackage{epstopdf}
\usepackage[usenames,dvipsnames,svgnames,table]{xcolor}
\usepackage{upgreek}
\begin{document}

\newcommand{\HaraldComment}[1]{\textcolor{red}{\textit{#1}}}
\newcommand{\JimComment}[1]{\textcolor{orange}{\textit{#1}}}

\newcommand{\degree}{^\circ}
\newcommand{\unit}[1]{\,\mathrm{#1}}
\newcommand{\Figureref}[1]{Figure~\ref{#1}}
\newcommand{\Eqnref}[1]{Equation~(\ref{#1})}
\newcommand{\bra}[1]{\left<\mathrm{#1}\right|}
\newcommand{\ket}[1]{\left|\mathrm{#1}\right>}

\title{Exploiting Rydberg Atom Surface Phonon Polariton Coupling for Single Photon Subtraction}

\author{H. K\"{u}bler}
\email[]{h.kuebler@physik.uni-stuttgart.de}
\affiliation{Homer L. Dodge Department of Physics and Astronomy, The University of Oklahoma, 440 W. Brooks St. Norman, OK 73019, USA}
\author{D. Booth}
\affiliation{Homer L. Dodge Department of Physics and Astronomy, The University of Oklahoma, 440 W. Brooks St. Norman, OK 73019, USA}
\author{J. Sedlacek}
\affiliation{Homer L. Dodge Department of Physics and Astronomy, The University of Oklahoma, 440 W. Brooks St. Norman, OK 73019, USA}
\author{P. Zabawa}
\affiliation{Homer L. Dodge Department of Physics and Astronomy, The University of Oklahoma, 440 W. Brooks St. Norman, OK 73019, USA}
\author{J.P. Shaffer}
\email[]{shaffer@nhn.ou.edu}
\affiliation{Homer L. Dodge Department of Physics and Astronomy, The University of Oklahoma, 440 W. Brooks St. Norman, OK 73019, USA}
\date{\today}

\begin{abstract}
  We investigate a hybrid quantum system that consists of a superatom coupled to a surface phonon-polariton. We apply this hybrid quantum system to subtract individual photons from a beam of light. Rydberg atom blockade is used to attain absorption of a single photon by an atomic microtrap. Surface phonon-polariton coupling to the superatom then triggers the transfer of the excitation to a storage state, a single Rydberg atom. The approach utilizes the interaction between a superatom and a Markovian bath that acts as a controlled decoherence mechanism to irreversibly project the superatom state into a single Rydberg atom state that can be read out.
\end{abstract}

\maketitle

Devices like quantum computers that rely on entanglement have proven difficult to realize and make robust. One school of thought suggests advances require linking quantum sub-systems that are individually tailored to meet specific challenges presented by effects such as dephasing, readout and interfacing to conventional electronics, so called hybrid quantum systems \cite{Girvin}. Consequently, developing experiments and theory for the useful interfacing of disparate quantum objects like atoms and surfaces is increasingly important and interesting.

\begin{figure}[t]
  \begin{minipage}[t]{0.48\columnwidth}
    \begin{flushleft}
      (a)
    \end{flushleft}
  \end{minipage}\hfill
  \begin{minipage}[t]{0.48\columnwidth}
    \begin{flushleft}
      (b)
    \end{flushleft}
  \end{minipage}\\
  \begin{minipage}{0.48\columnwidth}
    \includegraphics[width=\textwidth,trim=5cm 0cm 10cm 2.5cm, clip=true]{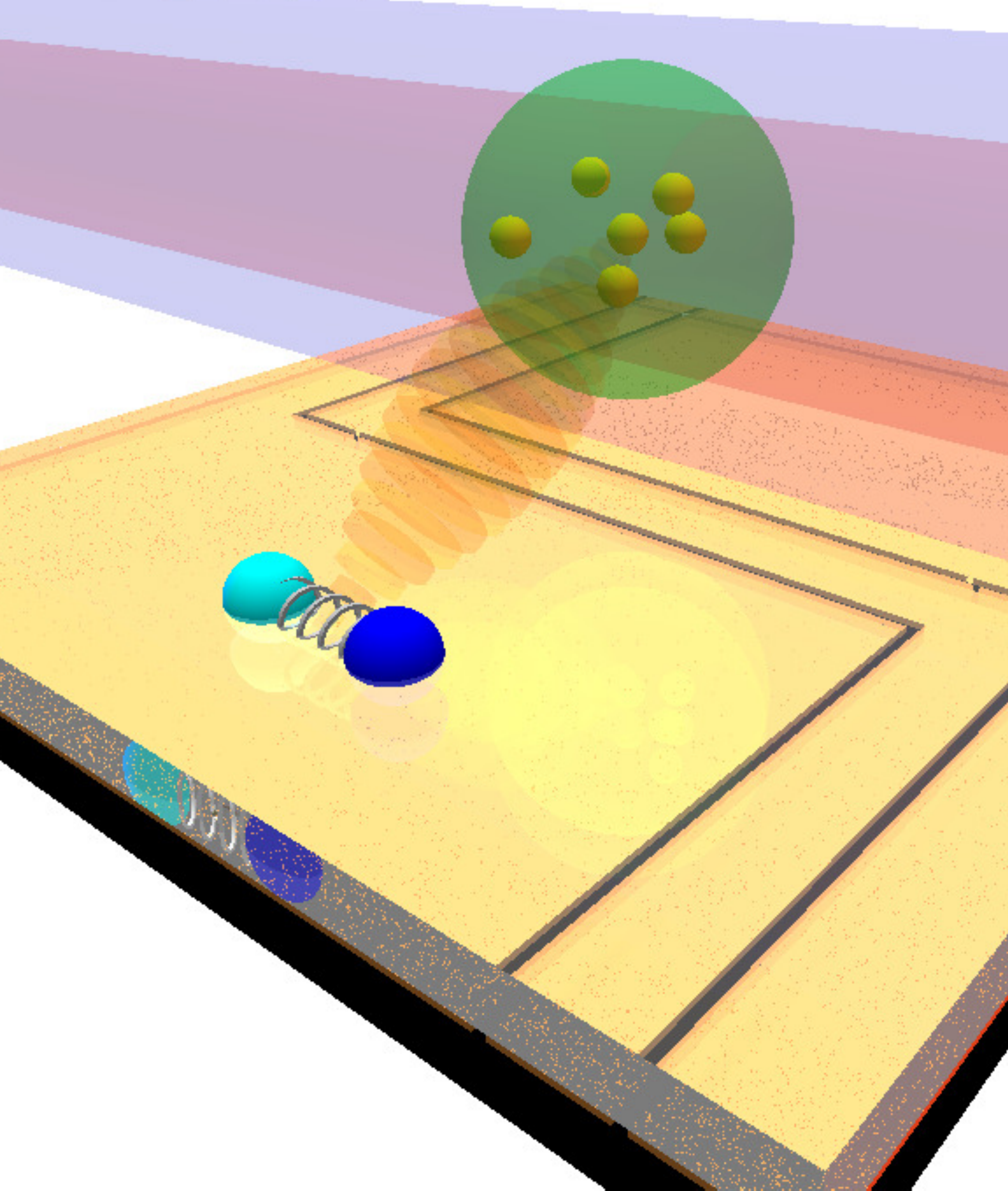}
  \end{minipage}\hfill
  \begin{minipage}{0.48\columnwidth}
    \includegraphics[width=\textwidth]{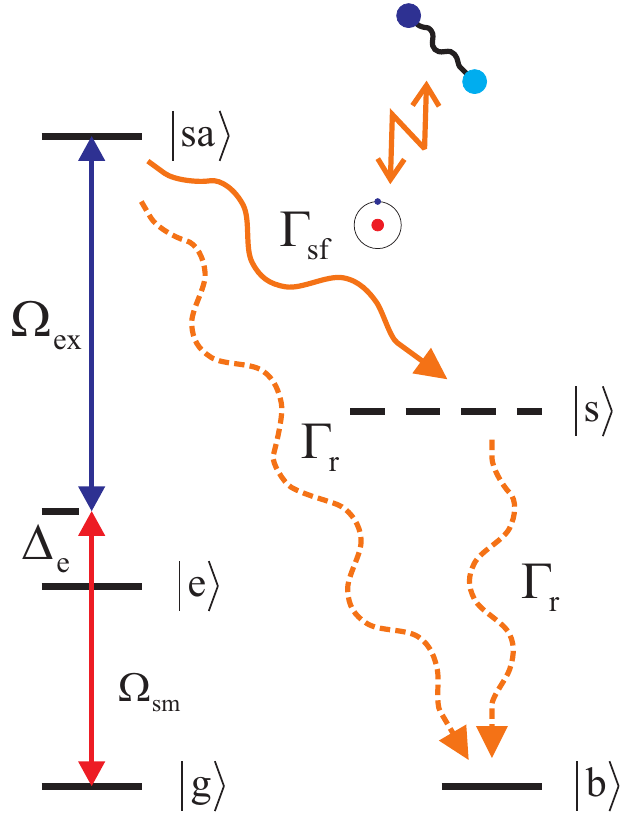}
  \end{minipage}

  \caption{ (Color online)
    (a) Schematic view of photon subtraction and counting concept. Atoms are trapped above an atom chip. On top of the chip is a dielectric substrate. Laser beams parallel to the chip surface excite atoms into a superatom state. The decay of the superatom is resonantly enhanced by coupling to a SPP, shown in the foreground. (b) Level scheme for single photon absorption. A two photon excitation, with coupling constants $\Omega_\mathrm{sm}$ and $\Omega_\mathrm{ex}$, produces a single excitation in a Rydberg state $\ket{a}$ shared by the trapped atoms. From there it decays with an enhanced rate $\Gamma_\mathrm{sf}$ due to the SPP coupling to $\ket{s}$. $\ket{b}$ represents all outcomes where information is lost, via decay of the Rydberg atoms at rate $\Gamma_\mathrm{r}$.
  }
  \label{fig:setup}
\end{figure}

In this paper, we investigate a hybrid quantum system that consists of a superatom coupled to a surface phonon-polariton (SPP) \cite{Shaffer11}, \Figureref{fig:setup}. A superatom is a single Rydberg excitation coherently shared by a cluster of atoms contained in a volume determined by the blockade radius, $r_\mathrm{b}$ \cite{Eyler04,Raithel05,Weidemuller04,HeidemannBlockade}. $r_\mathrm{b}$ is the distance over which pair interactions between Rydberg atoms \cite{ArneCalculationPaper,Cabral11} shift the excitation of a second Rydberg atom out of resonance with the laser driving the transition. The coupling of a superatom to a SPP is a resonant process because SPPs are discrete resonances that occur at the interface between 2 media when one has a negative dielectric constant and the other a positive one. Atom SPP coupling has been investigated previously in other contexts \cite{Hinds88,Ducloy02}.

We apply these concepts to design a scheme for subtracting individual photons from a beam of light, to either count photons or generate quantum light fields. We exploit the properties of Rydberg atom blockade to limit absorption by an atomic microtrap, whose size is smaller than $r_\mathrm{b}$, to a single photon \cite{SinglePhotonSubtractionBuechler}. After absorption, the single excitation is stored by coupling the superatom to a SPP that quickly decays into the bulk polariton modes of a dielectric. The coupling between the superatom and the SPP is resonantly enhanced so that a specific Rydberg atom storage state can be populated. The decay to the storage state is irreversible and decoheres the superatom, which is important for detecting the photon subtraction and decoupling the excitation from the light fields. The excitation phase of the process benefits from the $\sqrt{N}$ enhancement of the transition amplitude supplied by the superatom state, where $N$ is the number of atoms making up the superatom. In contrast, the storage state is a single excited Rydberg atom that decays at the rate of the populated Rydberg state. The dephasing of the superatom is accomplished via decay into the SPP and Rydberg storage state. The correlations of the SPP Markovian bath die away much faster than the coherent dynamics present in the overall system, effectively performing an irreversible measurement on the superatom. The Rydberg atom in the storage state blocks any further excitations. We show it is possible with current technologies to realize this scheme.

Most experiments and theory to date in quantum information focus on coherent coupling and explicitly try to avoid decoherence. However, in some cases, decoherence can aid in controlling and speeding up a desired quantum dynamical process \cite{NoisePlenio}. Controlled decoherence can, therefore, be a useful tool for designing quantum devices. One of the challenges of using decoherence as a tool is to introduce the noise in a controlled way. One possibility that has been realized is to use a laser speckle field \cite{SinglePhotonSubtractionBuechler,AndersonLocalisationBilly}, but this approach does not work for all applications. In our work, we introduce Rydberg atom SPP coupling \cite{MicroCellPaper} as a new, flexible and viable way to use decoherence in a controlled manner. Using the SPP coupling to a Rydberg atom allows dephasing to be controlled in many ways, including the distance dependence of the atom-SPP coupling, state selection and patterning of a surface with thin films to manipulate the SPP characteristics \cite{Agranovich,Sanders08}. More broadly, this hybrid quantum system offers the advantages of high frequency resonant coupling, the possibility of turning interactions on and off optically, terahertz coupling to conventional electronics and the ability to access ground atomic states with long coherence times. These properties can potentially be exploited to design other devices.

\section{General Idea}

For our device, we envision atoms being confined in a trap close to an atom chip, $\sim 3-10\unit{\upmu m}$, which is covered with a transparent dielectric, \Figureref{fig:setup}. The trap volume is smaller than the Rydberg atom blockade volume, $\sim 1 \unit{\upmu m^3}$. Traps with these dimensions have been realized experimentally \cite{MicrotrapSpreeuw,MicrotrapVideotape}. An electromagnetic field mode of a light source with Rabi frequency $\Omega_\mathrm{sm}$ is focused through the atomic cloud. Together with a strong excitation laser, $\Omega_\mathrm{ex}$, the source drives a detuned two-photon excitation to a Rydberg superatom state $\ket{sa}$, with $\Delta_\mathrm{e}\gg\Omega_\mathrm{ex}$, where $\Delta_\mathrm{e}$ is the detuning from an intermediate state $\ket{e}$. The single atom two photon Rabi frequency $\Omega_\mathrm{a}=\Omega_\mathrm{sm}\Omega_\mathrm{ex}/\left(4\Delta_\mathrm{e}\right)$ after adiabatically eliminationg $\ket{e}$. Under these conditions, the effective Rabi frequency for this superatom state is $\Omega_\mathrm{eff}=\sqrt{N}\Omega_\mathrm{a}$ \cite{HeidemannBlockade}. Since $\Delta_\mathrm{e}$ is chosen much larger than $\Omega_\mathrm{sm}$ and $\Omega_\mathrm{ex}$, the excitation linewidth is determined by $\Omega_\mathrm{eff}$ and the decay rate of the Rydberg state. When the Rydberg atom blockade shifts are much greater than the effective linewidth of the $\ket{g} \rightarrow \ket{sa}$ transition, all other photons from the source are transmitted through the trap.  A series of traps along the propagation direction of the source mode allows for multiple subtractions and counting of photons. The bandwidth of the device is determined by the range over which $\Omega_\mathrm{eff}$ can be large while still allowing the adiabatic elimination of $\ket{e}$ and the tuning range of the Rydberg excitation laser. The $\sqrt{N}$ enhancement in the excitation is the only place where the superatom concept is important. We use $\ket{a}$ to denote the Rydberg state used to form
\begin{equation}
  \ket{sa}=\frac{1}{\sqrt{N}}\sum_{i=1}^N\ket{g_1,g_2,\cdots,a_\mathit{i},\cdots, g_{\mathit{N}-1},g_\mathit{N}}.
\end{equation}

For a practical device, the superatom excitation needs to be irreversibly transferred to a storage state $\ket{s}$ so that it decouples from the light fields and can be read out \cite{SinglePhotonSubtractionBuechler}. In our scheme, the storage state $\ket{s}$ is one of the possible product states $\ket{s_\mathit{i}}=\ket{g_1,g_2,\cdots,s_\mathit{i},\cdots, g_{\mathit{N}-1},g_\mathit{N}}$.
It is important that the transfer process to $\ket{s}$ happens as fast as possible with maximum efficiency, implying optimization occurs for critical damping when considering the atom as an oscillator. For critical damping, the decay rate to $\ket{s}$ is two times faster than the effective two-photon Rabi frequency, $\Gamma_\mathrm{sf}=2\Omega_\mathrm{eff}$, since $\Omega_\mathrm{eff}$ is the analog of the classical oscillation frequency \cite{SinglePhotonSubtractionBuechler}. The transfer probability and time depend on the distance between the atom cloud and the surface providing a variable to tune them.

The excitation stored in $\ket{s}$ can be detected using the scheme described in \cite{BlockadeDetectionBuechlerZoller,BlockadeDetectionSaffmanMolmer,SinglePhotonSubtractionBuechler}. The remaining ground state atoms are used to detect whether there is an excitation in one of the $\ket{s_\mathit{i}}$. The excitation in $\ket{s_\mathit{i}}$ can be detected by setting up another EIT system with a different Rydberg state, $\ket{d}\neq\ket{a} \mathrm{or} \ket{s}$. The detection EIT is setup so the unshifted state $\ket{d}$ fulfills the two photon EIT resonance condition.  Under these conditions, the presence of $\ket{s}$, in the form of one of the $\ket{s_i}$, shifts $\ket{d}$ out of resonance changing the absorptive and refractive properties of the cloud \cite{EITFleischhauer,RydbergEITAshok}. The change in refractive index can be detected by means of a homodyne measurement. $\ket{d}$ should be chosen such that the blockade radius is close to the same size as $\ket{s}$ and $\ket{a}$. Similar to $\ket{s}$, $\ket{d}$ needs to be decoupled from the SPPs.

\section{Calculations}

To demonstrate our approach for realistic parameters, we chose $\ket{a}$ and $\ket{s}$ using several criteria. $\ket{a}$ is coupled via a two photon transition from $\ket{g}$, limiting us to $n\mathrm{S}$ and $n\mathrm{D}$ states for $\ket{a}$, assuming alkali atoms. The transition dipole moment between $\ket{a}$ and $\ket{s}$ has to be strong enough to achieve sufficient coupling between the Rydberg state and the SPP. The constraint on the SPP coupling limits $\ket{s}$ to $n'\mathrm{P}$ and $n'\mathrm{F}$ states, where $|n-n'|$ is small. The energy difference between $\ket{a}$ and $\ket{s}$ also has to be resonant with a SPP. For common materials, the frequency of the lowest SPP modes range from $\sim 40\unit{cm^{-1}}$($\approx 1.2\unit{THz}$) to $150\unit{cm^{-1}}$($\approx 4.5\unit{THz}$) and are discrete with widths ranging from $\sim 0.1-2\, \mathrm{cm}^{-1}$, e.g. $\mathrm{LaF_3}$: $41\unit{cm^{-1}}$\cite{LaF3Meltzer} and quartz: $128\unit{cm^{-1}}$\cite{SpitzerKleinmanQuartz}. $\ket{s}$ should also be chosen so it does not couple to SPPs, but provides a large enough blockade radius to prevent further light absorption by the cloud. Finally, $r_\mathrm{b}$ has to be large enough to provide full blockade for trap sizes on the order of $1\unit{\upmu m^3}$.

Given this set of constraints, we chose $^{87}\mathrm{Rb}$ with $\ket{a}=\ket{37S_{1/2}}$ for our calculations. In contrast to D-states with similar $n$, S-states provide repulsive Rydberg atom interactions, so no molecules can be formed \cite{Schwettmann07,Overstreet07,Overstreet09}. The transition energy to $\ket{s}=\ket{31P_{3/2}}$ is $\sim40.8\unit{cm^{-1}}$, matching the lowest SPP of $\mathrm{LaF_3}$. Another reason for choosing this transition is that $\mathrm{LaF_3}$ is commercially available as a single crystal that can be cut in different orientations and is easily polished so the surface is of high optical quality with surface variations of less than $1\,$nm. $\ket{a}$ and $\ket{s}$ have no other significant couplings to $\mathrm{LaF_3}$ SPPs. In general, the SPP modes of a material are tunable by changing the temperature and the orientation of the crystal surface. A $\sim40.8\unit{cm^{-1}}$ SPP has a wavelength of $244\unit{\upmu m}$ ensuring that the trap is in the near field regime for atom - surface separations of less than $\lambda/2 \pi \approx 39\unit{\upmu m}$. For our model calculations this gives $2 \pi z/\lambda \sim 0.1$, where $z$ is the distance between the atoms and the surface. These choices for the states reduce the effective system to the one shown in \Figureref{fig:setup}b.

\begin{figure}[t]
  \begin{flushleft}
    (a)\\
  \end{flushleft}
  \vspace{-20pt}
  \includegraphics[width=\columnwidth, trim=6.5mm 10mm 4.0mm 10mm, clip=true]{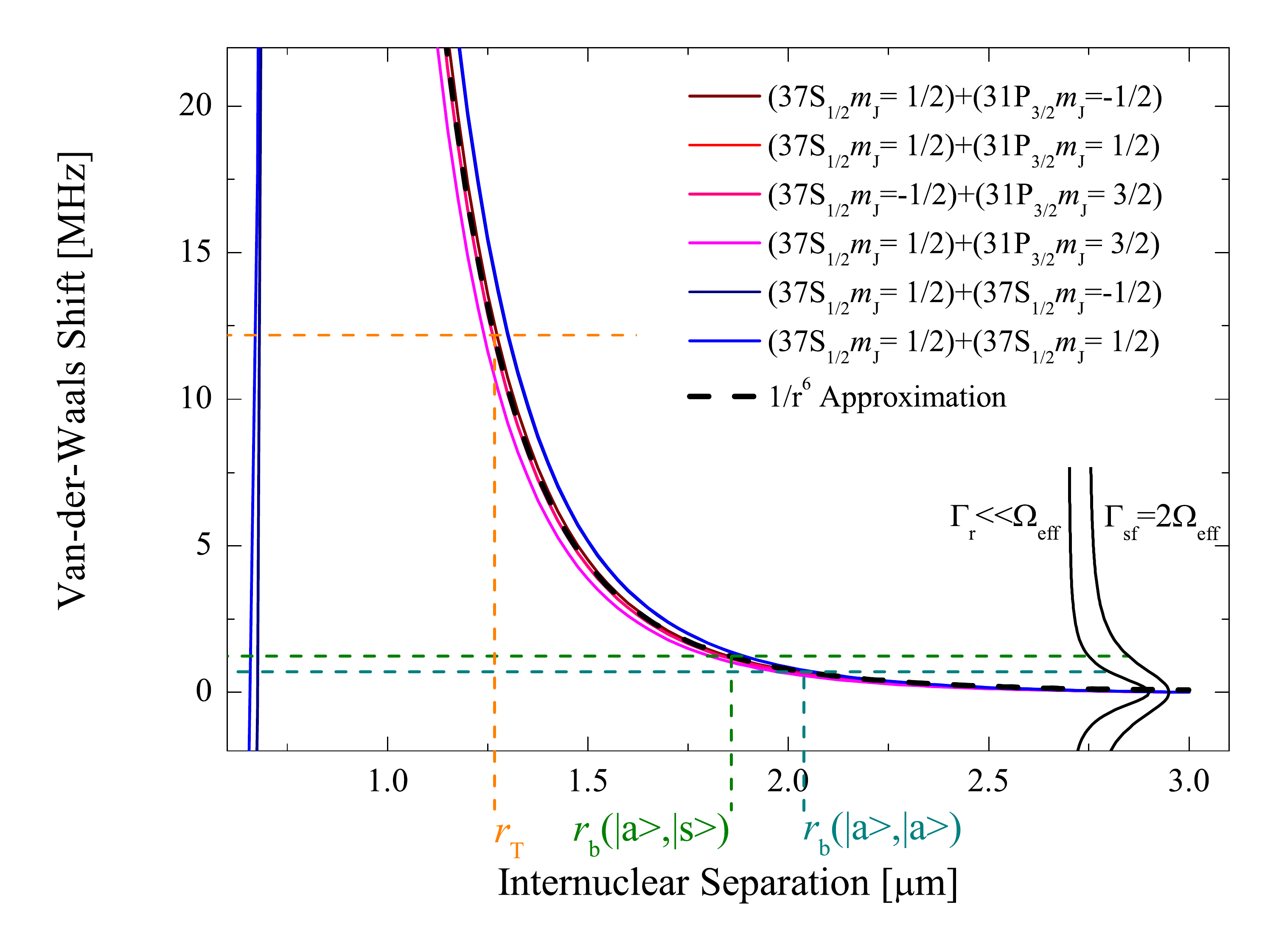}
  \begin{flushleft}
    (b)\\
  \end{flushleft}
  \vspace{-15pt}
  \includegraphics[width=\columnwidth, trim=0mm 0mm 0mm 0mm, clip=true]{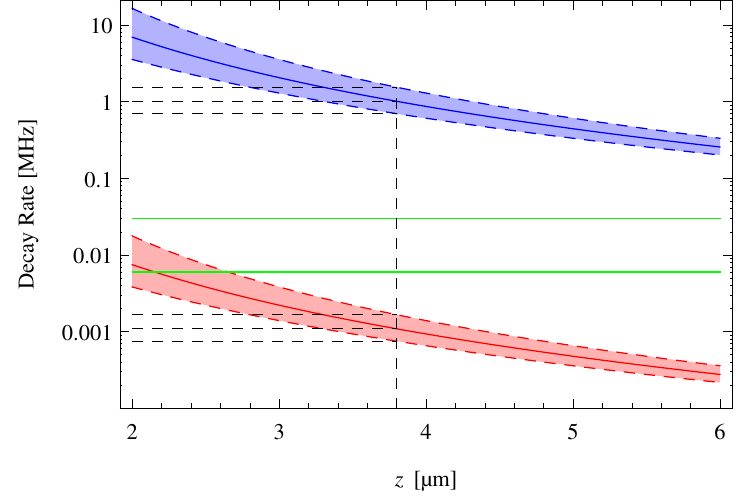}

  \caption{(Color online)
    (a) Rydberg pair interaction and blockade radius for the asymptotic states $(\ket{a},\ket{a})=(\ket{37S_{1/2}},\ket{37S_{1/2}})$ and $(\ket{a},\ket{s})=(\ket{37S_{1/2}},\ket{31P_{3/2}})$. For this combination of states the $C_6$ coefficients are similar. The excitation linewidth (dashed lines) for the case of strong $\Omega_\mathrm{eff}$ and critical damping are indicated, which defines blockade radii $r_\mathrm{b}$. $\Omega_\mathrm{eff}=2\pi\times1\unit{MHz}$. (b) $\mathrm{LaF_3}$ induced decay rate for state $\ket{a}$ (\textcolor{blue}{blue}) in resonance and (\textcolor{red}{red}) detuned by $3\unit{cm^{-1}}$ from a polariton resonance at $40.8\unit{cm^{-1}}$ with a width of $0.2 \unit{cm^{-1}}$ in a $1\unit{\upmu m}$ diameter trap. The dashed lines mark the spread due to the finite size of the trap. The proposed trapping distance of $3.8 \unit{\upmu m}$ is shown and the \textcolor{green}{green} lines indicate the decay rates $\Gamma_\mathrm{r}=6\unit{kHz}$ and $30\unit{kHz}$ used for simulations, see main text. The figure shows the superatom is weakly coupled to the surface if transitions are detuned by $> 3\unit{cm^{-1}}$. These transition rates decrease if the change in $n$ is increased, as the transition dipole is much reduced. Transitions to lower Rydberg states via other $\mathrm{LaF_3}$ SPP modes are not resonant, making them negligible.
  }
  \label{fig:interaction}
\end{figure}

\Figureref{fig:interaction}a shows a calculation of the energy shift, or Rydberg atom interaction, as a function of internuclear separation \cite{ArneCalculationPaper,Cabral11} for two atoms in $\ket{a}$ (red colors) and one in $\ket{a}$ and the other one in $\ket{s}$ (blue colors). The plots show that for  $\ket{a}=\ket{37S_{1/2}}$ and $\ket{s}=\ket{31P_{3/2}}$ the energy shift versus atom separation is almost the same for all relevant interactions. As a consequence, $r_\mathrm{b}$ does not change significantly if the excitation decays to $\ket{s}$. For the internuclear separations shown, the actual potentials are well approximated by a $\propto r^{-6}$ Van-der-Waals interaction.

On the lower right side of \Figureref{fig:interaction}a, the excitation linewidth for the cases when $\Omega_\mathrm{ex}$ dominates, and when the decay is optimized, $\Gamma_\mathrm{sf}=2\Omega_\mathrm{eff}$, are shown. For these calculations, an experimentally feasible $\Omega_\mathrm{eff}=2\pi\times1\unit{MHz}$ \cite{SaffmanRydbergBlockade} was used. By comparing these linewidths with the Van-der-Waals shift, $r_\mathrm{b}$ can be obtained. Once the shift is larger than the linewidth, the excitation is out of resonance with the light fields.
$r_\mathrm{b}$ without coupling to the polaritons is \cite{Eyler04,Raithel05,Weidemuller04,HeidemannBlockade}
\begin{equation}
  r_\mathrm{b}(\ket{a},\ket{a}) \approx
  \sqrt[6]{
    \frac{C_{6}(\ket{a},\ket{a})}
    {\hbar\sqrt{\Omega_\mathrm{eff}^2/2}}
  },
\end{equation}
but changes due to the shorter lifetime of $\ket{a}$ and the different $C_{6}$ coefficient to
\begin{equation}
  r_\mathrm{b}(\ket{a},\ket{s}) \approx
  \sqrt[6]{
    \frac{C_{6}(\ket{a},\ket{s})}
    {\hbar\sqrt{\left(\Gamma_\mathrm{sf}/2\right)^2+\Omega_\mathrm{eff}^2/2}}
  }.
\end{equation}
after decay to $\ket{s}$. The reduction of $r_\mathrm{b}$ due to optimization of the damping is as small as
$  r_\mathrm{b}(\ket{a},\ket{s})/r_\mathrm{b}(\ket{a},\ket{a}) \approx \sqrt[12]{1/3} \approx 0.9$ for $\Gamma_\mathrm{sf}=2\Omega_\mathrm{eff}$ and $C_{6}(\ket{a},\ket{a})\approx C_{6}(\ket{a},\ket{s})$.
To strongly suppress a second excitation in the trap, the blockade shift has to be much larger than the excitation linewidth. $r_\mathrm{T}\approx 1.25\unit{\upmu m}$ marks the distance at which the excitation probability drops below $1\%$. For further calculations $r_\mathrm{T}=1\unit{\upmu m}$ is used for the trap diameter. This argument and the plots in \Figureref{fig:interaction}a demonstrate that a Rydberg atom in $\ket{a}$ or $\ket{s}$ blocks the chance of more than one excitation occurring in a trap of size $r_\mathrm{T}=1\unit{\upmu m}$ with a probability $>99\%$.

To model the interaction between the Rydberg atom and SPP in the near field regime, we follow the approach in \cite{BartonPolaritonInteraction}. The transition dipole moment $\bra{a}\hat{d}\ket{s}$ couples the excitation in $\ket{a}$ to a SPP mode with frequency $\omega_\mathrm{pol}$ and linewidth $\Gamma_\mathrm{pol}$. $\Gamma_\mathrm{pol}$ results from decay of the SPP into bulk polariton modes. The Rydberg atom decay rate has a $z^{-3}$ dependence, where $z$ is the distance between the Rydberg atom and the surface:
\begin{equation}
  \Gamma_\mathrm{sf}=
    \frac{\sigma^2}{8\pi\epsilon_0 h z^3}
    \left|\bra{a}\hat{d}\ket{s}\right|^2
    \frac{\omega_\mathrm{pol}^2\omega_{\ket{a},\ket{s}}\Gamma_\mathrm{pol}}
      {\left(\omega_\mathrm{pol}^2-\omega_{\ket{a},\ket{s}}^2\right)^2+\omega_{\ket{a},\ket{s}}^2\Gamma_\mathrm{pol}^2},
      \label{eqn:damping}
\end{equation}
where $\sigma^2=(\varepsilon_0-1)/(\varepsilon_0+1)-(\varepsilon_\infty-1)/(\varepsilon_\infty+1)$ is the difference in polarizability of the dielectric at low and high frequency. $\omega_{\ket{a},\ket{s}}$ is the transition frequency between the states $\ket{a}$ and $\ket{s}$. The rate is enhanced at room temperature by a thermal factor $\Theta=(1-\exp{(-\beta\hbar\omega_\mathrm{pol})})^{-1}\approx 5.5$ \cite{BartonPolaritonInteraction}. If the SPP is resonant with the atomic transition, $\omega_{\ket{a},\ket{s}}=\omega_\mathrm{pol}$, the Lorentzian in \Eqnref{eqn:damping} reduces to a resonant factor $\omega_\mathrm{pol}/\Gamma_\mathrm{pol}$, which can be more than $100$. For example, the resonance in $\mathrm{LaF_3}$ (quartz) at $41\unit{cm^{-1}}$ ($394\unit{cm^{-1}}$) has a relative width $\gamma=0.005$ \cite{LaF3Meltzer} ($\gamma=0.007\pm0.001$ \cite{SpitzerKleinmanQuartz}) resulting in an resonant enhancement of $1/\gamma=200$ ($1/\gamma\approx 143$).
The resonant coupling rate between a single Rydberg atom and the SPP follows as
\begin{equation}
  \Gamma_\mathrm{sf,opt}=
    \frac{\sigma^2\Theta}{8\pi\epsilon_0 h z^3}
    \left|d_\mathrm{single}\right|^2
    \frac{\omega_\mathrm{pol}}{\Gamma_\mathrm{pol}}.
\end{equation}

$\Gamma_\mathrm{sf,opt}$ can be further increased by fabricating a thin metal film, or layers of such films, on the surface of the dielectric to create a SPP waveguide. A waveguide allows the SPPs to travel further on the surface by reducing their decay rate into the bulk polaritons. This can lead to a reduction in the SPP's linewidth by a factor $>10$ \cite{PolarisationNarrowing}, which increases $\Gamma_\mathrm{sf,opt}$ by the same factor. Other possibilities to increase the coupling are e.g. gratings \cite{PolaritonTayloringHasman}. An increase in the lifetime of the SPP's by a factor of $10$ does not change the fact that the decay of $\ket{sa}$ to $\ket{s}$ is irreversible as the decay of the SPP into bulk polaritons is still much faster than any other time scale present in the atomic dynamics.

This strong coupling between the SPPs and the bulk polariton modes of the dielectric leads to a fast transfer from the surface to the bulk, $\sim 1/\Gamma_{\mathrm{pol}}$. The bulk modes are a Markovian bath and all correlations die away on a timescale that is much faster than all the atomic dynamics. Due to this fast decoherence, the storage state is not a coherent superposition, but one of the single excited states $\ket{s_\mathit{i}}$.
The superatom can decay into each of the $\ket{s_\mathit{i}}$ equally, resulting in a total optimized decay rate
\begin{equation}
  \Gamma_\mathrm{sf,sa,opt}=\sum_{i=1}^N
    \frac{\sigma^2\Theta}{8\pi\epsilon_0 h z^3}
    \left|\bra{sa}\hat{d}\ket{s_\mathit{i}}\right|^2
    \frac{\omega_\mathrm{pol}}{\Gamma_\mathrm{pol}},
\end{equation}
where $\bra{sa}\hat{d}\ket{s_\mathit{i}}=d_\mathrm{single}/\sqrt{N}$. This decrease in the transition dipole moment is due to the reduced contribution of each $\ket{a_\mathit{i}}$ in $\ket{sa}$. Summing over the $N$ individual decay possibilities, each scaling as $d^2\propto 1/N$, results in a superaton decay rate $\Gamma_\mathrm{sf,sa,opt}=\Gamma_\mathrm{sf,opt}$, the single atom decay rate.

\Figureref{fig:interaction}b shows the resonant SPP decay rate (blue) for a superatom
as a function of $z$.
The single atom transition dipole moment $d_\mathrm{single}=\bra{37S_{1/2}}\hat{d}\ket{31P_{3/2}}\approx 15.5 \unit{ea_0}$ and $\Gamma_\mathrm{pol}=0.2\unit{cm^{-1}}$, corresponding to $\mathrm{LaF_3}$. The dielectric constants used for $\mathrm{LaF_3}$ are $\varepsilon_0=14$ and $\varepsilon_\infty=2.56$ \cite{LaF3Igel}. For a distance $z\approx3.8\unit{\upmu m}$, $\Gamma_{\mathrm{sf,opt}}=2\Omega_\mathrm{eff}\approx 2 \pi \times 2\unit{MHz}$ is achieved, sufficient to realize our scheme with a trap compatible with current methods \cite{MicrotrapSpreeuw,MicrotrapVideotape}.
Note that this is the distance to the surface of the dielectric, not to the atom chip itself.
Superatoms of this size can be realized in magnetic microtraps with sub-Poissonian number fluctuations at distance of $< 10\unit{\upmu m}$ \cite{MicrotrapNumberSpreeuw}.

After decay to $\ket{s}$, the coherence of the superatom is lost. If $\ket{s}$ is detuned from the same or another SPP by $15\Gamma_\mathrm{pol}$ (red) or more, its decay is completely determined by $\Gamma_{r,\mathrm{31P_{3/2}}}=2\pi \times 6.6\unit{kHz}$ \cite{RbRydbergLifetimeTheo,RbRydbergLifetimeEx}, the Rydberg atom decay rate, \Figureref{fig:interaction}b. A possible detection state $\ket{d}$, of which there are many possibilities, whose closest transition is also detuned by $15\Gamma_\mathrm{pol}$ (red), also couples to the surface at a rate comparable to $\Gamma_\mathrm{r}$ at these distances. \Figureref{fig:interaction}b shows that by choosing the other Rydberg states involved in the process so that they are detuned from SPP resonances by at least $3\unit{cm^{-1}}$, decay from those states via SPPs is negligible.

\begin{figure}[b]
    \includegraphics[width=\columnwidth, trim=0mm 0mm 0mm 0mm, clip=true]{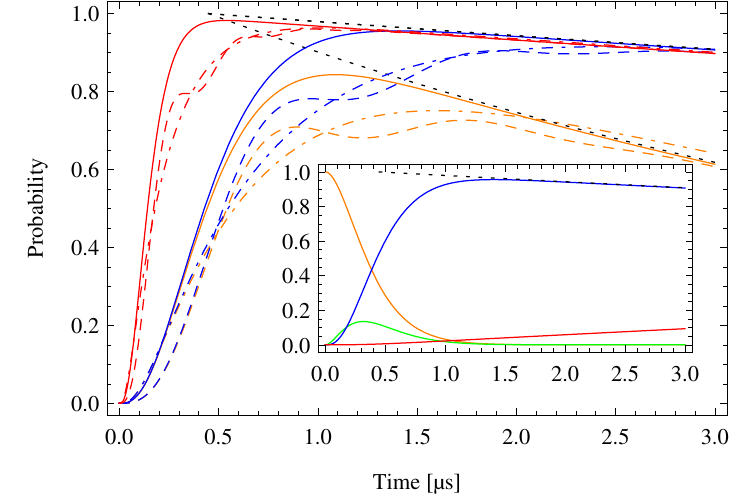}

  \caption{ (Color online)
    Probability of finding the system in $\ket{s}$ for different Rydberg atom decay rates $\Gamma_\mathrm{r}$ and effective Rabi frequencies $\Omega_\mathrm{eff}$. \textcolor{orange}{orange}: $\Gamma_\mathrm{r}=2\pi\times30\unit{kHz}$, $\Omega_\mathrm{eff}=2\pi\times1\unit{MHz}$, \textcolor{blue}{blue}: $\Gamma_\mathrm{r}=2\pi\times6\unit{kHz}$, $\Omega_\mathrm{eff}=2\pi\times1\unit{MHz}$, \textcolor{red}{red}: $\Gamma_\mathrm{r}=2\pi\times6\unit{kHz}$, $\Omega_\mathrm{eff}=2\pi\times3\unit{MHz}$. The dashed line represents the underdamped case, $\Gamma_\mathrm{sf}=\Omega_\mathrm{eff}/2$ and the dot-dashed line shows the overdamped regime, $\Gamma_\mathrm{sf}=4\Omega_\mathrm{eff}$. The solid line represents critical damping, $\Gamma_\mathrm{sf}=2\Omega_\mathrm{eff}$. The dotted black lines indicate the Rydberg decay, $\Gamma_\mathrm{r}$. Inset: Probability vs. time of finding the system in $\ket{g}$ (\textcolor{orange}{orange}), $\ket{a}$ (\textcolor{green}{green}), $\ket{s}$ (\textcolor{blue}{blue}) and $\ket{b}$ (\textcolor{red}{red}) for $\Omega_\mathrm{eff}=2\pi\times1\unit{MHz},$ $\Gamma_\mathrm{sf}=2\Omega,$ $\Gamma_\mathrm{r}=2\pi\times6\unit{kHz}$.
  }
  \label{fig:dyn}
\end{figure}

To simulate the system dynamics, we used a density matrix approach for the level scheme in \Figureref{fig:setup}b. We adiabatically eliminated $\ket{e}$ since $\Delta_\mathrm{e}\gg\Omega_\mathrm{ex}$ and used $\Omega_\mathrm{eff}$ to coherently couple $\ket{g}$ and $\ket{sa}$. The decay from $\ket{sa}$ into $\ket{s}$ via production of a SPP is modeled as an enhanced spontaneous decay with a rate $\Gamma_\mathrm{sf}$. Both Rydberg states $\ket{a}$ and $\ket{s}$ decay with their respective Rydberg decay rates $\Gamma_\mathrm{r}$ into $\ket{b}$, modeling the loss in the system dynamics.

\Figureref{fig:dyn} shows the result of the calculations for $\Omega_\mathrm{eff}=2\pi \times 1\unit{MHz}$ (blue). The solid line represents the dynamics for the optimized decay rate $\Gamma_\mathrm{sf}=2\Omega_\mathrm{eff}$. In this case, the transfer to $\ket{s}$ is the fastest and the readout time is only limited by the Rydberg decay. Our results indicate a readout window of more than $2\unit{\upmu s}$ is available with a fidelity $f>90\%$. The transfer is slower for the over- and underdamped regime (dot-dashed and dashed line), but $f = 90\%$ can still be achieved, albeit in a shorter time window. $f$ strongly depends on the Rydberg lifetime. Calculations for an increased decay rate are shown in \Figureref{fig:dyn} for $\Gamma_\mathrm{r}=2\pi \times 30\unit{kHz}$ instead of $2\pi \times 6\unit{kHz}$ (orange). In this case, the time window for $f>80\%$ is decreased to $< 1\unit{\upmu s}$. The performance can be improved with faster dynamics. The red lines in \Figureref{fig:dyn} show calculations for an effective Rabi frequency $\Omega_\mathrm{eff}=2\pi \times 3\unit{MHz}$ to demonstrate this point.

The inset in \Figureref{fig:dyn} shows the time evolution of the different states involved in the photon subtraction process for $\Gamma_\mathrm{sf}=2\Omega_\mathrm{eff}$. The Rydberg decay rate $\Gamma_\mathrm{r}=2\pi\times6\unit{kHz}$ for the plots in the figure. In the beginning, $\ket{g}$ (orange) is fully populated. Within the first $0.5\unit{\upmu s}$, the absorbing state $\ket{a}$ (green) is populated. However, due to the decay into $\ket{s}$, the population is very quickly transferred to the storage state $\ket{s}$ (blue). Only at longer times, $> 1\,\mu$s, does the Rydberg decay to the unwanted state $\ket{b}$ (red) become significant. The dashed line indicates this decay.
The calculations in \Figureref{fig:dyn} show that our approach is feasible and robust.

\section{Conclusion}

We have shown that Rydberg atom SPP coupling can be used to add decoherence in a controlled way to make a quantum device, a single photon counter or subtractor. The parameters discussed in the text are all experimentally achievable. The decoherence process can be sped up by trapping the atoms closer to the surface or utilizing stronger, perhaps optimally engineered \cite{PolaritonTayloringHasman}, Rydberg atom-SPP coupling. The SPP resonance can be narrowed by a factor $>10$ by coating the dielectric with thin metal films to form a waveguide \cite{PolarisationNarrowing}. A SPP waveguide can increase the Rydberg atom-SPP coupling and allows one to decrease the time required for the whole process. Likewise, patterning the surface in this manner can also enable one to increase the distance between the trap and the surface. Perhaps more significantly, we have introduced a new hybrid quantum system that can be further investigated for other quantum device applications. The Rydberg atom- SPP system is particularly interesting because of the spectral range where the couplings between the SPP and Rydberg atom lie, $\sim 1\unit{THz}$. We are currently investigating the couplings between Rydberg atoms and SPP's experimentally using optical measurement techniques we have developed for electric field measurement \cite{MWpaper,Sedlacek13}.

We thank G. Agarwal and T. Pfau for useful discussions. This work was supported by the NSF (PHY-1104424) and AFOSR (FA9550-12-1-0282).

\bibliography{mybib}

\begin{thebibliography}{37}%
\makeatletter
\providecommand \@ifxundefined [1]{%
 \@ifx{#1\undefined}
}%
\providecommand \@ifnum [1]{%
 \ifnum #1\expandafter \@firstoftwo
 \else \expandafter \@secondoftwo
 \fi
}%
\providecommand \@ifx [1]{%
 \ifx #1\expandafter \@firstoftwo
 \else \expandafter \@secondoftwo
 \fi
}%
\providecommand \natexlab [1]{#1}%
\providecommand \enquote  [1]{``#1''}%
\providecommand \bibnamefont  [1]{#1}%
\providecommand \bibfnamefont [1]{#1}%
\providecommand \citenamefont [1]{#1}%
\providecommand \href@noop [0]{\@secondoftwo}%
\providecommand \href [0]{\begingroup \@sanitize@url \@href}%
\providecommand \@href[1]{\@@startlink{#1}\@@href}%
\providecommand \@@href[1]{\endgroup#1\@@endlink}%
\providecommand \@sanitize@url [0]{\catcode `\\12\catcode `\$12\catcode
  `\&12\catcode `\#12\catcode `\^12\catcode `\_12\catcode `\%12\relax}%
\providecommand \@@startlink[1]{}%
\providecommand \@@endlink[0]{}%
\providecommand \url  [0]{\begingroup\@sanitize@url \@url }%
\providecommand \@url [1]{\endgroup\@href {#1}{\urlprefix }}%
\providecommand \urlprefix  [0]{URL }%
\providecommand \Eprint [0]{\href }%
\providecommand \doibase [0]{http://dx.doi.org/}%
\providecommand \selectlanguage [0]{\@gobble}%
\providecommand \bibinfo  [0]{\@secondoftwo}%
\providecommand \bibfield  [0]{\@secondoftwo}%
\providecommand \translation [1]{[#1]}%
\providecommand \BibitemOpen [0]{}%
\providecommand \bibitemStop [0]{}%
\providecommand \bibitemNoStop [0]{.\EOS\space}%
\providecommand \EOS [0]{\spacefactor3000\relax}%
\providecommand \BibitemShut  [1]{\csname bibitem#1\endcsname}%
\let\auto@bib@innerbib\@empty
\bibitem [{\citenamefont {Schoelkopf}\ and\ \citenamefont
  {Girvin}(2008)}]{Girvin}%
  \BibitemOpen
  \bibfield  {author} {\bibinfo {author} {\bibfnamefont {R.}~\bibnamefont
  {Schoelkopf}}\ and\ \bibinfo {author} {\bibfnamefont {S.}~\bibnamefont
  {Girvin}},\ }\href@noop {} {\bibfield  {journal} {\bibinfo  {journal}
  {Nature}\ }\textbf {\bibinfo {volume} {451}},\ \bibinfo {pages} {664}
  (\bibinfo {year} {2008})}\BibitemShut {NoStop}%
\bibitem [{\citenamefont {Shaffer}(2011)}]{Shaffer11}%
  \BibitemOpen
  \bibfield  {author} {\bibinfo {author} {\bibfnamefont {J.~P.}\ \bibnamefont
  {Shaffer}},\ }\href@noop {} {\bibfield  {journal} {\bibinfo  {journal} {Nat.
  Phot.}\ }\textbf {\bibinfo {volume} {5}},\ \bibinfo {pages} {451} (\bibinfo
  {year} {2011})}\BibitemShut {NoStop}%
\bibitem [{\citenamefont {Tong}\ \emph {et~al.}(2004)\citenamefont {Tong},
  \citenamefont {Farooqi}, \citenamefont {Stanojevic}, \citenamefont
  {Krishnan}, \citenamefont {Zhang}, \citenamefont {C\^ot\'e}, \citenamefont
  {Eyler},\ and\ \citenamefont {Gould}}]{Eyler04}%
  \BibitemOpen
  \bibfield  {author} {\bibinfo {author} {\bibfnamefont {D.}~\bibnamefont
  {Tong}}, \bibinfo {author} {\bibfnamefont {S.~M.}\ \bibnamefont {Farooqi}},
  \bibinfo {author} {\bibfnamefont {J.}~\bibnamefont {Stanojevic}}, \bibinfo
  {author} {\bibfnamefont {S.}~\bibnamefont {Krishnan}}, \bibinfo {author}
  {\bibfnamefont {Y.~P.}\ \bibnamefont {Zhang}}, \bibinfo {author}
  {\bibfnamefont {R.}~\bibnamefont {C\^ot\'e}}, \bibinfo {author}
  {\bibfnamefont {E.~E.}\ \bibnamefont {Eyler}}, \ and\ \bibinfo {author}
  {\bibfnamefont {P.~L.}\ \bibnamefont {Gould}},\ }\href {\doibase
  10.1103/PhysRevLett.93.063001} {\bibfield  {journal} {\bibinfo  {journal}
  {Phys. Rev. Lett.}\ }\textbf {\bibinfo {volume} {93}},\ \bibinfo {pages}
  {063001} (\bibinfo {year} {2004})}\BibitemShut {NoStop}%
\bibitem [{\citenamefont {Cubel-Liebisch}\ \emph {et~al.}(2005)\citenamefont
  {Cubel-Liebisch}, \citenamefont {Reinhard}, \citenamefont {Berman},\ and\
  \citenamefont {Raithel}}]{Raithel05}%
  \BibitemOpen
  \bibfield  {author} {\bibinfo {author} {\bibfnamefont {T.}~\bibnamefont
  {Cubel-Liebisch}}, \bibinfo {author} {\bibfnamefont {A.}~\bibnamefont
  {Reinhard}}, \bibinfo {author} {\bibfnamefont {P.~R.}\ \bibnamefont
  {Berman}}, \ and\ \bibinfo {author} {\bibfnamefont {G.}~\bibnamefont
  {Raithel}},\ }\href {\doibase 10.1103/PhysRevLett.95.253002} {\bibfield
  {journal} {\bibinfo  {journal} {Phys. Rev. Lett.}\ }\textbf {\bibinfo
  {volume} {95}},\ \bibinfo {pages} {253002} (\bibinfo {year}
  {2005})}\BibitemShut {NoStop}%
\bibitem [{\citenamefont {Singer}\ \emph {et~al.}(2004)\citenamefont {Singer},
  \citenamefont {Reetz-Lamour}, \citenamefont {Amthor}, \citenamefont
  {Marcassa},\ and\ \citenamefont {Weidem\"uller}}]{Weidemuller04}%
  \BibitemOpen
  \bibfield  {author} {\bibinfo {author} {\bibfnamefont {K.}~\bibnamefont
  {Singer}}, \bibinfo {author} {\bibfnamefont {M.}~\bibnamefont
  {Reetz-Lamour}}, \bibinfo {author} {\bibfnamefont {T.}~\bibnamefont
  {Amthor}}, \bibinfo {author} {\bibfnamefont {L.~G.}\ \bibnamefont
  {Marcassa}}, \ and\ \bibinfo {author} {\bibfnamefont {M.}~\bibnamefont
  {Weidem\"uller}},\ }\href {\doibase 10.1103/PhysRevLett.93.163001} {\bibfield
   {journal} {\bibinfo  {journal} {Phys. Rev. Lett.}\ }\textbf {\bibinfo
  {volume} {93}},\ \bibinfo {pages} {163001} (\bibinfo {year}
  {2004})}\BibitemShut {NoStop}%
\bibitem [{\citenamefont {Heidemann}\ \emph {et~al.}(2007)\citenamefont
  {Heidemann}, \citenamefont {Raitzsch}, \citenamefont {Bendkowsky},
  \citenamefont {Butscher}, \citenamefont {L\"ow}, \citenamefont {Santos},\
  and\ \citenamefont {Pfau}}]{HeidemannBlockade}%
  \BibitemOpen
  \bibfield  {author} {\bibinfo {author} {\bibfnamefont {R.}~\bibnamefont
  {Heidemann}}, \bibinfo {author} {\bibfnamefont {U.}~\bibnamefont {Raitzsch}},
  \bibinfo {author} {\bibfnamefont {V.}~\bibnamefont {Bendkowsky}}, \bibinfo
  {author} {\bibfnamefont {B.}~\bibnamefont {Butscher}}, \bibinfo {author}
  {\bibfnamefont {R.}~\bibnamefont {L\"ow}}, \bibinfo {author} {\bibfnamefont
  {L.}~\bibnamefont {Santos}}, \ and\ \bibinfo {author} {\bibfnamefont
  {T.}~\bibnamefont {Pfau}},\ }\href {\doibase 10.1103/PhysRevLett.99.163601}
  {\bibfield  {journal} {\bibinfo  {journal} {Phys. Rev. Lett.}\ }\textbf
  {\bibinfo {volume} {99}},\ \bibinfo {pages} {163601} (\bibinfo {year}
  {2007})}\BibitemShut {NoStop}%
\bibitem [{\citenamefont {Schwettmann}\ \emph {et~al.}(2006)\citenamefont
  {Schwettmann}, \citenamefont {Crawford}, \citenamefont {Overstreet},\ and\
  \citenamefont {Shaffer}}]{ArneCalculationPaper}%
  \BibitemOpen
  \bibfield  {author} {\bibinfo {author} {\bibfnamefont {A.}~\bibnamefont
  {Schwettmann}}, \bibinfo {author} {\bibfnamefont {J.}~\bibnamefont
  {Crawford}}, \bibinfo {author} {\bibfnamefont {K.~R.}\ \bibnamefont
  {Overstreet}}, \ and\ \bibinfo {author} {\bibfnamefont {J.~P.}\ \bibnamefont
  {Shaffer}},\ }\href {\doibase 10.1103/PhysRevA.74.020701} {\bibfield
  {journal} {\bibinfo  {journal} {Phys. Rev. A}\ }\textbf {\bibinfo {volume}
  {74}},\ \bibinfo {pages} {020701} (\bibinfo {year} {2006})}\BibitemShut
  {NoStop}%
\bibitem [{\citenamefont {Cabral}\ \emph {et~al.}(2011)\citenamefont {Cabral},
  \citenamefont {Kondo}, \citenamefont {Gon\c{c}alves}, \citenamefont
  {Nascimento}, \citenamefont {Marcassa}, \citenamefont {Booth}, \citenamefont
  {Tallant}, \citenamefont {Schwettmann}, \citenamefont {Overstreet},
  \citenamefont {Sedlacek},\ and\ \citenamefont {Shaffer}}]{Cabral11}%
  \BibitemOpen
  \bibfield  {author} {\bibinfo {author} {\bibfnamefont {J.~S.}\ \bibnamefont
  {Cabral}}, \bibinfo {author} {\bibfnamefont {J.~M.}\ \bibnamefont {Kondo}},
  \bibinfo {author} {\bibfnamefont {L.~F.}\ \bibnamefont {Gon\c{c}alves}},
  \bibinfo {author} {\bibfnamefont {V.~A.}\ \bibnamefont {Nascimento}},
  \bibinfo {author} {\bibfnamefont {L.~G.}\ \bibnamefont {Marcassa}}, \bibinfo
  {author} {\bibfnamefont {D.}~\bibnamefont {Booth}}, \bibinfo {author}
  {\bibfnamefont {J.}~\bibnamefont {Tallant}}, \bibinfo {author} {\bibfnamefont
  {A.}~\bibnamefont {Schwettmann}}, \bibinfo {author} {\bibfnamefont {K.~R.}\
  \bibnamefont {Overstreet}}, \bibinfo {author} {\bibfnamefont
  {J.}~\bibnamefont {Sedlacek}}, \ and\ \bibinfo {author} {\bibfnamefont
  {J.~P.}\ \bibnamefont {Shaffer}},\ }\href@noop {} {\bibfield  {journal}
  {\bibinfo  {journal} {J. Phys. B}\ }\textbf {\bibinfo {volume} {44}},\
  \bibinfo {pages} {184007} (\bibinfo {year} {2011})}\BibitemShut {NoStop}%
\bibitem [{\citenamefont {Anderson}\ \emph {et~al.}(1988)\citenamefont
  {Anderson}, \citenamefont {Haroche}, \citenamefont {Hinds}, \citenamefont
  {Jhe},\ and\ \citenamefont {Meschede}}]{Hinds88}%
  \BibitemOpen
  \bibfield  {author} {\bibinfo {author} {\bibfnamefont {A.}~\bibnamefont
  {Anderson}}, \bibinfo {author} {\bibfnamefont {S.}~\bibnamefont {Haroche}},
  \bibinfo {author} {\bibfnamefont {E.~A.}\ \bibnamefont {Hinds}}, \bibinfo
  {author} {\bibfnamefont {W.}~\bibnamefont {Jhe}}, \ and\ \bibinfo {author}
  {\bibfnamefont {D.}~\bibnamefont {Meschede}},\ }\href {\doibase
  10.1103/PhysRevA.37.3594} {\bibfield  {journal} {\bibinfo  {journal} {Phys.
  Rev. A}\ }\textbf {\bibinfo {volume} {37}},\ \bibinfo {pages} {3594}
  (\bibinfo {year} {1988})}\BibitemShut {NoStop}%
\bibitem [{\citenamefont {Failache}\ \emph {et~al.}(2002)\citenamefont
  {Failache}, \citenamefont {Saltiel}, \citenamefont {Fischer}, \citenamefont
  {Bloch},\ and\ \citenamefont {Ducloy}}]{Ducloy02}%
  \BibitemOpen
  \bibfield  {author} {\bibinfo {author} {\bibfnamefont {H.}~\bibnamefont
  {Failache}}, \bibinfo {author} {\bibfnamefont {S.}~\bibnamefont {Saltiel}},
  \bibinfo {author} {\bibfnamefont {A.}~\bibnamefont {Fischer}}, \bibinfo
  {author} {\bibfnamefont {D.}~\bibnamefont {Bloch}}, \ and\ \bibinfo {author}
  {\bibfnamefont {M.}~\bibnamefont {Ducloy}},\ }\href {\doibase
  10.1103/PhysRevLett.88.243603} {\bibfield  {journal} {\bibinfo  {journal}
  {Phys. Rev. Lett.}\ }\textbf {\bibinfo {volume} {88}},\ \bibinfo {pages}
  {243603} (\bibinfo {year} {2002})}\BibitemShut {NoStop}%
\bibitem [{\citenamefont {Honer}\ \emph {et~al.}(2011)\citenamefont {Honer},
  \citenamefont {L\"ow}, \citenamefont {Weimer}, \citenamefont {Pfau},\ and\
  \citenamefont {B\"uchler}}]{SinglePhotonSubtractionBuechler}%
  \BibitemOpen
  \bibfield  {author} {\bibinfo {author} {\bibfnamefont {J.}~\bibnamefont
  {Honer}}, \bibinfo {author} {\bibfnamefont {R.}~\bibnamefont {L\"ow}},
  \bibinfo {author} {\bibfnamefont {H.}~\bibnamefont {Weimer}}, \bibinfo
  {author} {\bibfnamefont {T.}~\bibnamefont {Pfau}}, \ and\ \bibinfo {author}
  {\bibfnamefont {H.~P.}\ \bibnamefont {B\"uchler}},\ }\href {\doibase
  10.1103/PhysRevLett.107.093601} {\bibfield  {journal} {\bibinfo  {journal}
  {Phys. Rev. Lett.}\ }\textbf {\bibinfo {volume} {107}},\ \bibinfo {pages}
  {093601} (\bibinfo {year} {2011})}\BibitemShut {NoStop}%
\bibitem [{\citenamefont {Plenio}\ and\ \citenamefont
  {F.}(2008)}]{NoisePlenio}%
  \BibitemOpen
  \bibfield  {author} {\bibinfo {author} {\bibfnamefont {M.~B.}\ \bibnamefont
  {Plenio}}\ and\ \bibinfo {author} {\bibfnamefont {H.~S.}\ \bibnamefont
  {F.}},\ }\href {http://stacks.iop.org/1367-2630/10/i=11/a=113019} {\bibfield
  {journal} {\bibinfo  {journal} {New J. Phys.}\ }\textbf {\bibinfo {volume}
  {10}},\ \bibinfo {pages} {113019} (\bibinfo {year} {2008})}\BibitemShut
  {NoStop}%
\bibitem [{\citenamefont {Billy}\ \emph {et~al.}(2008)\citenamefont {Billy},
  \citenamefont {Josse}, \citenamefont {Zuo}, \citenamefont {Bernard},
  \citenamefont {Hambrecht}, \citenamefont {Lugan}, \citenamefont {Clement},
  \citenamefont {Sanchez-Palencia}, \citenamefont {Bouyer},\ and\ \citenamefont
  {Aspect}}]{AndersonLocalisationBilly}%
  \BibitemOpen
  \bibfield  {author} {\bibinfo {author} {\bibfnamefont {J.}~\bibnamefont
  {Billy}}, \bibinfo {author} {\bibfnamefont {V.}~\bibnamefont {Josse}},
  \bibinfo {author} {\bibfnamefont {Z.}~\bibnamefont {Zuo}}, \bibinfo {author}
  {\bibfnamefont {A.}~\bibnamefont {Bernard}}, \bibinfo {author} {\bibfnamefont
  {B.}~\bibnamefont {Hambrecht}}, \bibinfo {author} {\bibfnamefont
  {P.}~\bibnamefont {Lugan}}, \bibinfo {author} {\bibfnamefont
  {D.}~\bibnamefont {Clement}}, \bibinfo {author} {\bibfnamefont
  {L.}~\bibnamefont {Sanchez-Palencia}}, \bibinfo {author} {\bibfnamefont
  {P.}~\bibnamefont {Bouyer}}, \ and\ \bibinfo {author} {\bibfnamefont
  {A.}~\bibnamefont {Aspect}},\ }\href {\doibase 10.1038/nature07000}
  {\bibfield  {journal} {\bibinfo  {journal} {Nature}\ }\textbf {\bibinfo
  {volume} {453}},\ \bibinfo {pages} {891} (\bibinfo {year}
  {2008})}\BibitemShut {NoStop}%
\bibitem [{\citenamefont {K\"ubler}\ \emph {et~al.}(2010)\citenamefont
  {K\"ubler}, \citenamefont {Shaffer}, \citenamefont {Baluktsian},
  \citenamefont {L\"ow},\ and\ \citenamefont {Pfau}}]{MicroCellPaper}%
  \BibitemOpen
  \bibfield  {author} {\bibinfo {author} {\bibfnamefont {H.}~\bibnamefont
  {K\"ubler}}, \bibinfo {author} {\bibfnamefont {J.}~\bibnamefont {Shaffer}},
  \bibinfo {author} {\bibfnamefont {T.}~\bibnamefont {Baluktsian}}, \bibinfo
  {author} {\bibfnamefont {R.}~\bibnamefont {L\"ow}}, \ and\ \bibinfo {author}
  {\bibfnamefont {T.}~\bibnamefont {Pfau}},\ }\href
  {http://dx.doi.org/10.1038/nphoton.2009.260} {\bibfield  {journal} {\bibinfo
  {journal} {Nat. Phot.}\ }\textbf {\bibinfo {volume} {4}},\ \bibinfo {pages}
  {112} (\bibinfo {year} {2010})}\BibitemShut {NoStop}%
\bibitem [{\citenamefont {Agranovich}\ and\ \citenamefont
  {Mills}(1982)}]{Agranovich}%
  \BibitemOpen
  \bibfield  {author} {\bibinfo {author} {\bibfnamefont {V.}~\bibnamefont
  {Agranovich}}\ and\ \bibinfo {author} {\bibfnamefont {D.}~\bibnamefont
  {Mills}},\ }\href@noop {} {\emph {\bibinfo {title} {Surface Polaritons:
  Electromagnetic Waves at Surfaces and Interfaces, Volume 1}}},\ Modern
  Problems in Condensed Matter Sciences\ (\bibinfo  {publisher} {North
  Holland},\ \bibinfo {year} {1982})\BibitemShut {NoStop}%
\bibitem [{\citenamefont {Kamli}\ \emph {et~al.}(2008)\citenamefont {Kamli},
  \citenamefont {Moiseev},\ and\ \citenamefont {Sanders}}]{Sanders08}%
  \BibitemOpen
  \bibfield  {author} {\bibinfo {author} {\bibfnamefont {A.}~\bibnamefont
  {Kamli}}, \bibinfo {author} {\bibfnamefont {S.~A.}\ \bibnamefont {Moiseev}},
  \ and\ \bibinfo {author} {\bibfnamefont {B.~C.}\ \bibnamefont {Sanders}},\
  }\href {\doibase 10.1103/PhysRevLett.101.263601} {\bibfield  {journal}
  {\bibinfo  {journal} {Phys. Rev. Lett.}\ }\textbf {\bibinfo {volume} {101}},\
  \bibinfo {pages} {263601} (\bibinfo {year} {2008})}\BibitemShut {NoStop}%
\bibitem [{\citenamefont {Leung}\ \emph {et~al.}(2011)\citenamefont {Leung},
  \citenamefont {Tauschinsky}, \citenamefont {Druten},\ and\ \citenamefont
  {Spreeuw}}]{MicrotrapSpreeuw}%
  \BibitemOpen
  \bibfield  {author} {\bibinfo {author} {\bibfnamefont {V.}~\bibnamefont
  {Leung}}, \bibinfo {author} {\bibfnamefont {A.}~\bibnamefont {Tauschinsky}},
  \bibinfo {author} {\bibfnamefont {N.}~\bibnamefont {Druten}}, \ and\ \bibinfo
  {author} {\bibfnamefont {R.}~\bibnamefont {Spreeuw}},\ }\href {\doibase
  10.1007/s11128-011-0295-1} {\bibfield  {journal} {\bibinfo  {journal} {Quant.
  Inf. Proc.}\ }\textbf {\bibinfo {volume} {10}},\ \bibinfo {pages} {955}
  (\bibinfo {year} {2011})}\BibitemShut {NoStop}%
\bibitem [{\citenamefont {Sinclair}\ \emph {et~al.}(2005)\citenamefont
  {Sinclair}, \citenamefont {Retter}, \citenamefont {Curtis}, \citenamefont
  {Hall}, \citenamefont {Llorente~Garcia}, \citenamefont {Eriksson},
  \citenamefont {Sauer},\ and\ \citenamefont {Hinds}}]{MicrotrapVideotape}%
  \BibitemOpen
  \bibfield  {author} {\bibinfo {author} {\bibfnamefont {C.~D.}\ \bibnamefont
  {Sinclair}}, \bibinfo {author} {\bibfnamefont {J.~A.}\ \bibnamefont
  {Retter}}, \bibinfo {author} {\bibfnamefont {E.~A.}\ \bibnamefont {Curtis}},
  \bibinfo {author} {\bibfnamefont {B.~V.}\ \bibnamefont {Hall}}, \bibinfo
  {author} {\bibfnamefont {I.}~\bibnamefont {Llorente~Garcia}}, \bibinfo
  {author} {\bibfnamefont {S.}~\bibnamefont {Eriksson}}, \bibinfo {author}
  {\bibfnamefont {B.~E.}\ \bibnamefont {Sauer}}, \ and\ \bibinfo {author}
  {\bibfnamefont {E.~A.}\ \bibnamefont {Hinds}},\ }\href {\doibase
  10.1140/epjd/e2005-00088-6} {\bibfield  {journal} {\bibinfo  {journal} {Eur.
  Phys. D}\ }\textbf {\bibinfo {volume} {35}},\ \bibinfo {pages} {105}
  (\bibinfo {year} {2005})}\BibitemShut {NoStop}%
\bibitem [{\citenamefont {M\"uller}\ \emph {et~al.}(2009)\citenamefont
  {M\"uller}, \citenamefont {Lesanovsky}, \citenamefont {Weimer}, \citenamefont
  {B\"uchler},\ and\ \citenamefont {Zoller}}]{BlockadeDetectionBuechlerZoller}%
  \BibitemOpen
  \bibfield  {author} {\bibinfo {author} {\bibfnamefont {M.}~\bibnamefont
  {M\"uller}}, \bibinfo {author} {\bibfnamefont {I.}~\bibnamefont
  {Lesanovsky}}, \bibinfo {author} {\bibfnamefont {H.}~\bibnamefont {Weimer}},
  \bibinfo {author} {\bibfnamefont {H.~P.}\ \bibnamefont {B\"uchler}}, \ and\
  \bibinfo {author} {\bibfnamefont {P.}~\bibnamefont {Zoller}},\ }\href
  {\doibase 10.1103/PhysRevLett.102.170502} {\bibfield  {journal} {\bibinfo
  {journal} {Phys. Rev. Lett.}\ }\textbf {\bibinfo {volume} {102}},\ \bibinfo
  {pages} {170502} (\bibinfo {year} {2009})}\BibitemShut {NoStop}%
\bibitem [{\citenamefont {Saffman}\ and\ \citenamefont
  {M\o{}lmer}(2009)}]{BlockadeDetectionSaffmanMolmer}%
  \BibitemOpen
  \bibfield  {author} {\bibinfo {author} {\bibfnamefont {M.}~\bibnamefont
  {Saffman}}\ and\ \bibinfo {author} {\bibfnamefont {K.}~\bibnamefont
  {M\o{}lmer}},\ }\href {\doibase 10.1103/PhysRevLett.102.240502} {\bibfield
  {journal} {\bibinfo  {journal} {Phys. Rev. Lett.}\ }\textbf {\bibinfo
  {volume} {102}},\ \bibinfo {pages} {240502} (\bibinfo {year}
  {2009})}\BibitemShut {NoStop}%
\bibitem [{\citenamefont {Fleischhauer}\ \emph {et~al.}(2005)\citenamefont
  {Fleischhauer}, \citenamefont {Imamoglu},\ and\ \citenamefont
  {Marangos}}]{EITFleischhauer}%
  \BibitemOpen
  \bibfield  {author} {\bibinfo {author} {\bibfnamefont {M.}~\bibnamefont
  {Fleischhauer}}, \bibinfo {author} {\bibfnamefont {A.}~\bibnamefont
  {Imamoglu}}, \ and\ \bibinfo {author} {\bibfnamefont {J.~P.}\ \bibnamefont
  {Marangos}},\ }\href {\doibase 10.1103/RevModPhys.77.633} {\bibfield
  {journal} {\bibinfo  {journal} {Rev. Mod. Phys.}\ }\textbf {\bibinfo {volume}
  {77}},\ \bibinfo {pages} {633} (\bibinfo {year} {2005})}\BibitemShut
  {NoStop}%
\bibitem [{\citenamefont {Mohapatra}\ \emph {et~al.}(2007)\citenamefont
  {Mohapatra}, \citenamefont {Jackson},\ and\ \citenamefont
  {Adams}}]{RydbergEITAshok}%
  \BibitemOpen
  \bibfield  {author} {\bibinfo {author} {\bibfnamefont {A.~K.}\ \bibnamefont
  {Mohapatra}}, \bibinfo {author} {\bibfnamefont {T.~R.}\ \bibnamefont
  {Jackson}}, \ and\ \bibinfo {author} {\bibfnamefont {C.~S.}\ \bibnamefont
  {Adams}},\ }\href {\doibase 10.1103/PhysRevLett.98.113003} {\bibfield
  {journal} {\bibinfo  {journal} {Phys. Rev. Lett.}\ }\textbf {\bibinfo
  {volume} {98}},\ \bibinfo {pages} {113003} (\bibinfo {year}
  {2007})}\BibitemShut {NoStop}%
\bibitem [{\citenamefont {Meltzer}\ \emph {et~al.}(1983)\citenamefont
  {Meltzer}, \citenamefont {Rives},\ and\ \citenamefont {Dixon}}]{LaF3Meltzer}%
  \BibitemOpen
  \bibfield  {author} {\bibinfo {author} {\bibfnamefont {R.~S.}\ \bibnamefont
  {Meltzer}}, \bibinfo {author} {\bibfnamefont {J.~E.}\ \bibnamefont {Rives}},
  \ and\ \bibinfo {author} {\bibfnamefont {G.~S.}\ \bibnamefont {Dixon}},\
  }\href {\doibase 10.1103/PhysRevB.28.4786} {\bibfield  {journal} {\bibinfo
  {journal} {Phys. Rev. B}\ }\textbf {\bibinfo {volume} {28}},\ \bibinfo
  {pages} {4786} (\bibinfo {year} {1983})}\BibitemShut {NoStop}%
\bibitem [{\citenamefont {Spitzer}\ and\ \citenamefont
  {Kleinman}(1961)}]{SpitzerKleinmanQuartz}%
  \BibitemOpen
  \bibfield  {author} {\bibinfo {author} {\bibfnamefont {W.~G.}\ \bibnamefont
  {Spitzer}}\ and\ \bibinfo {author} {\bibfnamefont {D.~A.}\ \bibnamefont
  {Kleinman}},\ }\href {\doibase 10.1103/PhysRev.121.1324} {\bibfield
  {journal} {\bibinfo  {journal} {Phys. Rev.}\ }\textbf {\bibinfo {volume}
  {121}},\ \bibinfo {pages} {1324} (\bibinfo {year} {1961})}\BibitemShut
  {NoStop}%
\bibitem [{\citenamefont {Schwettmann}\ \emph {et~al.}(2007)\citenamefont
  {Schwettmann}, \citenamefont {Overstreet}, \citenamefont {Tallant},\ and\
  \citenamefont {Shaffer}}]{Schwettmann07}%
  \BibitemOpen
  \bibfield  {author} {\bibinfo {author} {\bibfnamefont {A.}~\bibnamefont
  {Schwettmann}}, \bibinfo {author} {\bibfnamefont {K.~R.}\ \bibnamefont
  {Overstreet}}, \bibinfo {author} {\bibfnamefont {J.}~\bibnamefont {Tallant}},
  \ and\ \bibinfo {author} {\bibfnamefont {J.~P.}\ \bibnamefont {Shaffer}},\
  }\href {\doibase 10.1080/09500340701584076} {\bibfield  {journal} {\bibinfo
  {journal} {J. Mod. Opt.}\ }\textbf {\bibinfo {volume} {54}},\ \bibinfo
  {pages} {2551} (\bibinfo {year} {2007})}\BibitemShut {NoStop}%
\bibitem [{\citenamefont {Overstreet}\ \emph {et~al.}(2007)\citenamefont
  {Overstreet}, \citenamefont {Schwettmann}, \citenamefont {Tallant},\ and\
  \citenamefont {Shaffer}}]{Overstreet07}%
  \BibitemOpen
  \bibfield  {author} {\bibinfo {author} {\bibfnamefont {K.~R.}\ \bibnamefont
  {Overstreet}}, \bibinfo {author} {\bibfnamefont {A.}~\bibnamefont
  {Schwettmann}}, \bibinfo {author} {\bibfnamefont {J.}~\bibnamefont
  {Tallant}}, \ and\ \bibinfo {author} {\bibfnamefont {J.~P.}\ \bibnamefont
  {Shaffer}},\ }\href {\doibase 10.1103/PhysRevA.76.011403} {\bibfield
  {journal} {\bibinfo  {journal} {Phys. Rev. A}\ }\textbf {\bibinfo {volume}
  {76}},\ \bibinfo {pages} {011403} (\bibinfo {year} {2007})}\BibitemShut
  {NoStop}%
\bibitem [{\citenamefont {Overstreet}\ \emph {et~al.}(2009)\citenamefont
  {Overstreet}, \citenamefont {Schwettmann}, \citenamefont {Tallant},
  \citenamefont {Booth},\ and\ \citenamefont {Shaffer}}]{Overstreet09}%
  \BibitemOpen
  \bibfield  {author} {\bibinfo {author} {\bibfnamefont {K.~R.}\ \bibnamefont
  {Overstreet}}, \bibinfo {author} {\bibfnamefont {A.}~\bibnamefont
  {Schwettmann}}, \bibinfo {author} {\bibfnamefont {J.}~\bibnamefont
  {Tallant}}, \bibinfo {author} {\bibfnamefont {D.}~\bibnamefont {Booth}}, \
  and\ \bibinfo {author} {\bibfnamefont {J.~P.}\ \bibnamefont {Shaffer}},\
  }\href@noop {} {\bibfield  {journal} {\bibinfo  {journal} {Nat. Phys.}\
  }\textbf {\bibinfo {volume} {5}},\ \bibinfo {pages} {581} (\bibinfo {year}
  {2009})}\BibitemShut {NoStop}%
\bibitem [{\citenamefont {Urban}\ \emph {et~al.}(2009)\citenamefont {Urban},
  \citenamefont {Johnson}, \citenamefont {Henage}, \citenamefont {Isenhower},
  \citenamefont {Yavuz}, \citenamefont {Walker},\ and\ \citenamefont
  {Saffman}}]{SaffmanRydbergBlockade}%
  \BibitemOpen
  \bibfield  {author} {\bibinfo {author} {\bibfnamefont {E.}~\bibnamefont
  {Urban}}, \bibinfo {author} {\bibfnamefont {T.~A.}\ \bibnamefont {Johnson}},
  \bibinfo {author} {\bibfnamefont {T.}~\bibnamefont {Henage}}, \bibinfo
  {author} {\bibfnamefont {L.}~\bibnamefont {Isenhower}}, \bibinfo {author}
  {\bibfnamefont {D.~D.}\ \bibnamefont {Yavuz}}, \bibinfo {author}
  {\bibfnamefont {T.~G.}\ \bibnamefont {Walker}}, \ and\ \bibinfo {author}
  {\bibfnamefont {M.}~\bibnamefont {Saffman}},\ }\href {\doibase
  10.1038/nphys1178} {\bibfield  {journal} {\bibinfo  {journal} {Nat. Phys.}\
  }\textbf {\bibinfo {volume} {5}},\ \bibinfo {pages} {110} (\bibinfo {year}
  {2009})}\BibitemShut {NoStop}%
\bibitem [{\citenamefont {Barton}(1997)}]{BartonPolaritonInteraction}%
  \BibitemOpen
  \bibfield  {author} {\bibinfo {author} {\bibfnamefont {G.}~\bibnamefont
  {Barton}},\ }\href {\doibase 10.1098/rspa.1997.0132} {\bibfield  {journal}
  {\bibinfo  {journal} {Proc. R. Soc. A}\ }\textbf {\bibinfo {volume} {453}},\
  \bibinfo {pages} {2461} (\bibinfo {year} {1997})}\BibitemShut {NoStop}%
\bibitem [{\citenamefont {Adato}\ and\ \citenamefont
  {Guo}(2007)}]{PolarisationNarrowing}%
  \BibitemOpen
  \bibfield  {author} {\bibinfo {author} {\bibfnamefont {R.}~\bibnamefont
  {Adato}}\ and\ \bibinfo {author} {\bibfnamefont {J.}~\bibnamefont {Guo}},\
  }\href {\doibase 10.1117/12.733335} {\enquote {\bibinfo {title} {Novel
  metal-dielectric structures for guiding ultra-long-range surface
  plasmon-polaritons at optical frequencies},}\ } (\bibinfo {year}
  {2007})\BibitemShut {NoStop}%
\bibitem [{\citenamefont {Dahan}\ \emph {et~al.}(2005)\citenamefont {Dahan},
  \citenamefont {Niv}, \citenamefont {Biener}, \citenamefont {Kleiner},\ and\
  \citenamefont {Hasman}}]{PolaritonTayloringHasman}%
  \BibitemOpen
  \bibfield  {author} {\bibinfo {author} {\bibfnamefont {N.}~\bibnamefont
  {Dahan}}, \bibinfo {author} {\bibfnamefont {A.}~\bibnamefont {Niv}}, \bibinfo
  {author} {\bibfnamefont {G.}~\bibnamefont {Biener}}, \bibinfo {author}
  {\bibfnamefont {V.}~\bibnamefont {Kleiner}}, \ and\ \bibinfo {author}
  {\bibfnamefont {E.}~\bibnamefont {Hasman}},\ }\href {\doibase
  10.1063/1.1922084} {\bibfield  {journal} {\bibinfo  {journal} {Appl. Phys.
  Lett.}\ }\textbf {\bibinfo {volume} {86}},\ \bibinfo {eid} {191102} (\bibinfo
  {year} {2005})}\BibitemShut {NoStop}%
\bibitem [{\citenamefont {Igel}\ \emph {et~al.}(1982)\citenamefont {Igel},
  \citenamefont {Wintersgill}, \citenamefont {Fontanella}, \citenamefont
  {Chadwick}, \citenamefont {Andeen},\ and\ \citenamefont {Bean}}]{LaF3Igel}%
  \BibitemOpen
  \bibfield  {author} {\bibinfo {author} {\bibfnamefont {J.~R.}\ \bibnamefont
  {Igel}}, \bibinfo {author} {\bibfnamefont {M.~C.}\ \bibnamefont
  {Wintersgill}}, \bibinfo {author} {\bibfnamefont {J.~J.}\ \bibnamefont
  {Fontanella}}, \bibinfo {author} {\bibfnamefont {A.~V.}\ \bibnamefont
  {Chadwick}}, \bibinfo {author} {\bibfnamefont {C.~G.}\ \bibnamefont
  {Andeen}}, \ and\ \bibinfo {author} {\bibfnamefont {V.~E.}\ \bibnamefont
  {Bean}},\ }\href {http://stacks.iop.org/0022-3719/15/i=35/a=019} {\bibfield
  {journal} {\bibinfo  {journal} {J. Phys. C}\ }\textbf {\bibinfo {volume}
  {15}},\ \bibinfo {pages} {7215} (\bibinfo {year} {1982})}\BibitemShut
  {NoStop}%
\bibitem [{\citenamefont {Whitlock}\ \emph {et~al.}(2010)\citenamefont
  {Whitlock}, \citenamefont {Ockeloen},\ and\ \citenamefont
  {Spreeuw}}]{MicrotrapNumberSpreeuw}%
  \BibitemOpen
  \bibfield  {author} {\bibinfo {author} {\bibfnamefont {S.}~\bibnamefont
  {Whitlock}}, \bibinfo {author} {\bibfnamefont {C.~F.}\ \bibnamefont
  {Ockeloen}}, \ and\ \bibinfo {author} {\bibfnamefont {R.~J.~C.}\ \bibnamefont
  {Spreeuw}},\ }\href {\doibase 10.1103/PhysRevLett.104.120402} {\bibfield
  {journal} {\bibinfo  {journal} {Phys. Rev. Lett.}\ }\textbf {\bibinfo
  {volume} {104}},\ \bibinfo {pages} {120402} (\bibinfo {year}
  {2010})}\BibitemShut {NoStop}%
\bibitem [{\citenamefont {Beterov}\ \emph {et~al.}(2009)\citenamefont
  {Beterov}, \citenamefont {Ryabtsev}, \citenamefont {Tretyakov},\ and\
  \citenamefont {Entin}}]{RbRydbergLifetimeTheo}%
  \BibitemOpen
  \bibfield  {author} {\bibinfo {author} {\bibfnamefont {I.~I.}\ \bibnamefont
  {Beterov}}, \bibinfo {author} {\bibfnamefont {I.~I.}\ \bibnamefont
  {Ryabtsev}}, \bibinfo {author} {\bibfnamefont {D.~B.}\ \bibnamefont
  {Tretyakov}}, \ and\ \bibinfo {author} {\bibfnamefont {V.~M.}\ \bibnamefont
  {Entin}},\ }\href {\doibase 10.1103/PhysRevA.79.052504} {\bibfield  {journal}
  {\bibinfo  {journal} {Phys. Rev. A}\ }\textbf {\bibinfo {volume} {79}},\
  \bibinfo {pages} {052504} (\bibinfo {year} {2009})}\BibitemShut {NoStop}%
\bibitem [{\citenamefont {Branden}\ \emph {et~al.}(2010)\citenamefont
  {Branden}, \citenamefont {Juhasz}, \citenamefont {Mahlokozera}, \citenamefont
  {Vesa}, \citenamefont {Wilson}, \citenamefont {Zheng}, \citenamefont
  {Kortyna},\ and\ \citenamefont {Tate}}]{RbRydbergLifetimeEx}%
  \BibitemOpen
  \bibfield  {author} {\bibinfo {author} {\bibfnamefont {D.~B.}\ \bibnamefont
  {Branden}}, \bibinfo {author} {\bibfnamefont {T.}~\bibnamefont {Juhasz}},
  \bibinfo {author} {\bibfnamefont {T.}~\bibnamefont {Mahlokozera}}, \bibinfo
  {author} {\bibfnamefont {C.}~\bibnamefont {Vesa}}, \bibinfo {author}
  {\bibfnamefont {R.~O.}\ \bibnamefont {Wilson}}, \bibinfo {author}
  {\bibfnamefont {M.}~\bibnamefont {Zheng}}, \bibinfo {author} {\bibfnamefont
  {A.}~\bibnamefont {Kortyna}}, \ and\ \bibinfo {author} {\bibfnamefont
  {D.~A.}\ \bibnamefont {Tate}},\ }\href
  {http://stacks.iop.org/0953-4075/43/i=1/a=015002} {\bibfield  {journal}
  {\bibinfo  {journal} {J. Phys. B}\ }\textbf {\bibinfo {volume} {43}},\
  \bibinfo {pages} {015002} (\bibinfo {year} {2010})}\BibitemShut {NoStop}%
\bibitem [{\citenamefont {Sedlacek}\ \emph {et~al.}(2012)\citenamefont
  {Sedlacek}, \citenamefont {Schwettmann}, \citenamefont {K\"ubler},
  \citenamefont {L\"ow}, \citenamefont {Pfau},\ and\ \citenamefont
  {Shaffer}}]{MWpaper}%
  \BibitemOpen
  \bibfield  {author} {\bibinfo {author} {\bibfnamefont {J.~A.}\ \bibnamefont
  {Sedlacek}}, \bibinfo {author} {\bibfnamefont {A.}~\bibnamefont
  {Schwettmann}}, \bibinfo {author} {\bibfnamefont {H.}~\bibnamefont
  {K\"ubler}}, \bibinfo {author} {\bibfnamefont {R.}~\bibnamefont {L\"ow}},
  \bibinfo {author} {\bibfnamefont {T.}~\bibnamefont {Pfau}}, \ and\ \bibinfo
  {author} {\bibfnamefont {J.~P.}\ \bibnamefont {Shaffer}},\ }\href@noop {}
  {\bibfield  {journal} {\bibinfo  {journal} {Nat. Phys.}\ }\textbf {\bibinfo
  {volume} {8}},\ \bibinfo {pages} {819} (\bibinfo {year} {2012})}\BibitemShut
  {NoStop}%
\bibitem [{\citenamefont {Sedlacek}\ \emph {et~al.}(2013)\citenamefont
  {Sedlacek}, \citenamefont {Schwettmann}, \citenamefont {K\"ubler},\ and\
  \citenamefont {Shaffer}}]{Sedlacek13}%
  \BibitemOpen
  \bibfield  {author} {\bibinfo {author} {\bibfnamefont {J.~A.}\ \bibnamefont
  {Sedlacek}}, \bibinfo {author} {\bibfnamefont {A.}~\bibnamefont
  {Schwettmann}}, \bibinfo {author} {\bibfnamefont {H.}~\bibnamefont
  {K\"ubler}}, \ and\ \bibinfo {author} {\bibfnamefont {J.~P.}\ \bibnamefont
  {Shaffer}},\ }\href {\doibase 10.1103/PhysRevLett.111.063001} {\bibfield
  {journal} {\bibinfo  {journal} {Phys. Rev. Lett.}\ }\textbf {\bibinfo
  {volume} {111}},\ \bibinfo {pages} {063001} (\bibinfo {year}
  {2013})}\BibitemShut {NoStop}%
\end{thebibliography}%

\end{document}